\documentclass[11pt,twoside]{article}

\usepackage{asp2004}
\usepackage{epsf}
\usepackage{psfig}
\usepackage{lscape}
\usepackage{graphicx}

\newcommand{\kms}{\ensuremath{{\rm km\,sec^{-1}}}}                   
\newcommand{\msun}{\ensuremath{\mathit{M}_{\odot}}}                  
\newcommand{\msunyr}{\ensuremath{\mathit{M}_{\odot}{\rm yr}^{-1}}}   
\newcommand{\lsun}{\ensuremath{\mathit{L}_{\odot}}}                  

\newcommand{\logg}{\ensuremath{\log \mathrm{g}}}                     
\newcommand{\lstar}{\ensuremath{\mathit{L}_{\star}}}                 
\newcommand{\mdot}{\ensuremath{\dot{M}}}                             
\newcommand{\teff}{\ensuremath{\mathit{T}_{\rm eff}}}                
\newcommand{\vinf}{\ensuremath{v_{\infty}}}                          
\newcommand{\vesc}{\ensuremath{v_{\rm esc}}}                         

\begin{document}
\title{Properties of Galactic B supergiants}    
\author{Paul A. Crowther}

\affil{Dept of Physics \& Astronomy, University of Sheffield, Hounsfield Rd, She
ffield, S3 7RH,  UK}

\author{Daniel J. Lennon}

\affil{Isaac Newton Group, Apartado 321, 38700 Santa Cruz de La Palma, Canary Islands, 
Spain}

\author{Nolan R. Walborn}

\affil{Space Telescope Science Institute, 3700 San Martin Drive, Baltimore MD 21218, USA}

\author{Stephen J. Smartt}   

\affil{Department of Physics and Astronomy, Queen's University Belfast, Belfast BT7 1NN, UK}    

\begin{abstract} Physical and wind properties of Galactic B supergiants
are presented based upon non-LTE line blanketed model atmospheres,
including Sher~25 toward the NGC~3603 cluster. We compare H$\alpha$
derived wind densities with recent results for SMC B supergiants and
generally confirm theoretical expectations for stronger winds amongst
Galactic supergiants. Mid B supergiant winds are substantially weaker than
predictions from current radiatively driven wind theory, a problem which
is exacerbated if winds are already clumped in the H$\alpha$ line forming
region. We find that the so-called `bistability jump' at B1 (\teff\ $\sim$
21kK) from Lamers et al. is rather a more gradual downward trend.  CNO
elemental abundances, including Sher~25, reveal partially processed
material at their surfaces. In general, these are in good agreement with
evolutionary predictions for blue supergiants evolving redward accounting
for rotational mixing. A few cases, including HD~152236 ($\zeta^{1}$ Sco),
 exhibit strongly processed material which is 
more typical of Luminous Blue Variables. Our  derived 
photospheric [N/O] ratio for Sher~25 agrees with
that for its ring nebula, although a higher degree of CNO processing would
be expected if the nebula originated during a red supergiant phase, as is
suspected for the ring nebula ejected by the B supergiant progenitor of
SN~1987A, Sk--69$^{\circ}$ 202. Sher~25 has an inferred age of $\sim$5Myr in
contrast with $\sim$2Myr for HD~97950, the ionizing cluster of NGC~3603,
so it may be a foreground object or close binary evolution may be responsible
for its unusual location in the H-R diagram.
\end{abstract}


\section{Introduction}

O stars dominate the energetics of young starbursts since their ionizing
output is a very steep function of effective temperature (Leitherer et al.
1992). As such, numerous intensive studies of O stars in the Milky Way
and Magellanic Clouds have been undertaken, with regard to determining
their fundamental parameters and the empirical dependence of their
wind strengths on metallicity (e.g. Mokiem et al. 2006ab).  
Unfortunately, O dwarfs are visually faint in external galaxies due to
their large bolometric corrections and small radii, with typically $M_{\rm 
V}$ = -5.0 mag.   O star abundances of CNO elements are difficult to 
derive (Crowther et al. 2002), yet these are sensitive to early rotational 
mixing (e.g. Meynet \& Maeder 2000).

In contrast, B supergiants have received rather less attention since they
are less relevant for energetics of young starbursts. Nevertheless, they
are visually bright, typically $M_{\rm V} = -7.0$ mag, possess larger
radii and so are easy to observe individually in nearby galaxies,
permitting robust tests of the metallicity dependence of radiatively
driven wind theory, which is predicted to increase from early to mid B
supergiants (Vink et al. 2000). In addition, CNO abundances are readily
obtained from optical spectroscopy, for comparison to evolutionary model
predictions. Notably, Trundle et al. (2004) and Trundle \& Lennon (2005)
have studied a large sample of SMC B supergiants using the line blanketed
version of FASTWIND (Puls et al. 2005). To date, the most extensive study
of Galactic B supergiants is by Kudritzki et al. (1999), who adopted the
\teff\ scale from the plane-parallel unblanketed study of McErlean et al.
(1999), plus mass-loss rates from the unblanketed version of FASTWIND.

Our present study of Galactic B supergiants employs CMFGEN (e.g.  Hillier
et al. 2003) which also incorporates line blanketing and spherical
geometry, and so permits a direct comparison with recent SMC results, plus
CNO abundance determinations. Do the abundances of normal B
supergiants match those of stars evolving towards red supergiants (RSG) or
returning from RSG following convective dredge-up? Comparisons 
with Luminous Blue Variables (LBVs) are also made, since they
exhibit spectral characteristics of early B supergiants at
visual minimum and A supergiants at visual maximum (Crowther 1997).  
Notably, $\zeta^{1}$ Sco (HD~152236) is an early B hypergiant and lies
above the empirical Humphreys-Davidson limit (Humphreys \& Davidson 1979). 
We also include new stellar and nebular studies of Sher~25
(Hendry et al. 2006), an apparently normal B supergiant in the giant HII
region NGC~3603 yet possessing an ejecta nebula reminiscent of that
associated with SN~1987A (Brandner et al.  1997b).

Finally, we consider the empirical bistability jump amongst early B
supergiants, which was originally developed for the LBV P~Cygni by
Pauldrach \& Puls (1990), in the sense that slight variations in its
effective temperature resulted in hydrogen being ionized or neutral in its
outer stellar wind (Najarro et al.  1997).  Lamers et al. (1995) studied
wind velocities of OB stars, and established a discontinuity in the ratio
of \vinf\ to \vesc\ for B0.5 and B1 supergiants, which is assumed in
current stellar wind models of Vink et al. (2000).

\begin{figure}[ht!] \begin{center}
\includegraphics[width=0.6\columnwidth,clip,angle=-90]{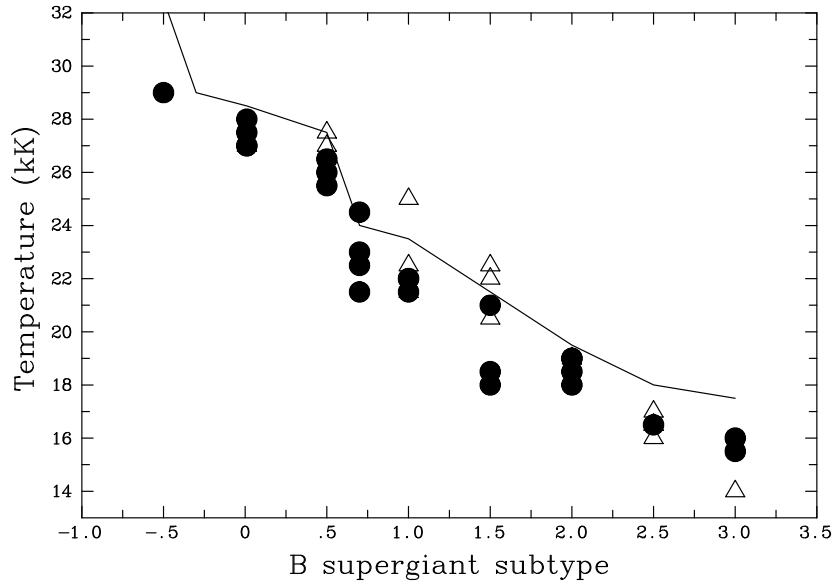}
\end{center} \caption{Temperature scale of B supergiants from this Milky
Way study (CMFGEN, circles) plus Trundle et al. (2004) and Trundle \&
Lennon (2005) for the SMC (FASTWIND, triangles) versus the McErlean et al.
(1998, solid line) calibration.\label{teff}} \end{figure}

\section{Observations and analysis}

Full details of observations and method of analysis are 
presented by Crowther et al. (2006). In summary
we have observed 26 Galactic early-type Ia supergiants -- spanning the
spectral range O9.5 to B3 -- at intermediate dispersion in the blue and
red using the 1.0m JHK, 2.5m INT or 4.2m WHT for the northern sample or
at the 1.5m CTIO for the southern sample, with the exception of Sher~25 
for which 3.9m AAT echelle spectroscopy was obtained. Cluster or 
association
distance estimates are available for 75\% of the sample, whilst average
subtype absolute magnitudes are taken from Magellanic Cloud stars for the 
remainder. For the case of Sher~25, we adopt a distance of 7.9~kpc 
(Russeil 2003) towards
NGC~3603 from which we obtain $M_{\rm V}=-7.4$ mag. 

For the present study we have used CMFGEN (see Hillier et al. 2003) which
solves the radiative transfer equation in the co-moving frame, under the
constraints of radiative and statistical equilibrium. For the supersonic
part, the velocity is parameterized with a classical $\beta$ law, which is
connected to a hydrostatic density structure at depth using a H-He TLUSTY
model. A depth independent Doppler profile was assumed for the atmospheric
structure calculation in the co-moving frame, whilst a radially dependent
turbulence was used to calculate the emergent spectrum in the observer's
frame, in which incoherent electron scattering and Stark broadening for H,
He were included.

Stellar temperatures were derived from suitable lines of HeI-II (O9.5-B0),
SiIII--IV (B0.5--B2), SiII-III (B2.5-B3) such that radii and luminosities
followed from absolute magnitudes, with H$\alpha$ observations providing
\mdot\ and the velocity exponent $\beta$. Terminal velocities and
rotational velocities were taken from Howarth et al. (1997). H/He
abundances are difficult to accurately derive in B supergiants, such that
He/H=0.2 by number is adopted throughout. CNO abundances were varied to
reproduce the relevant optical absorption lines, and compared with Solar
abundances (Asplund et al. 2004).

\begin{figure}[ht!] \begin{center}
\includegraphics[width=0.7\columnwidth,clip,angle=-90]{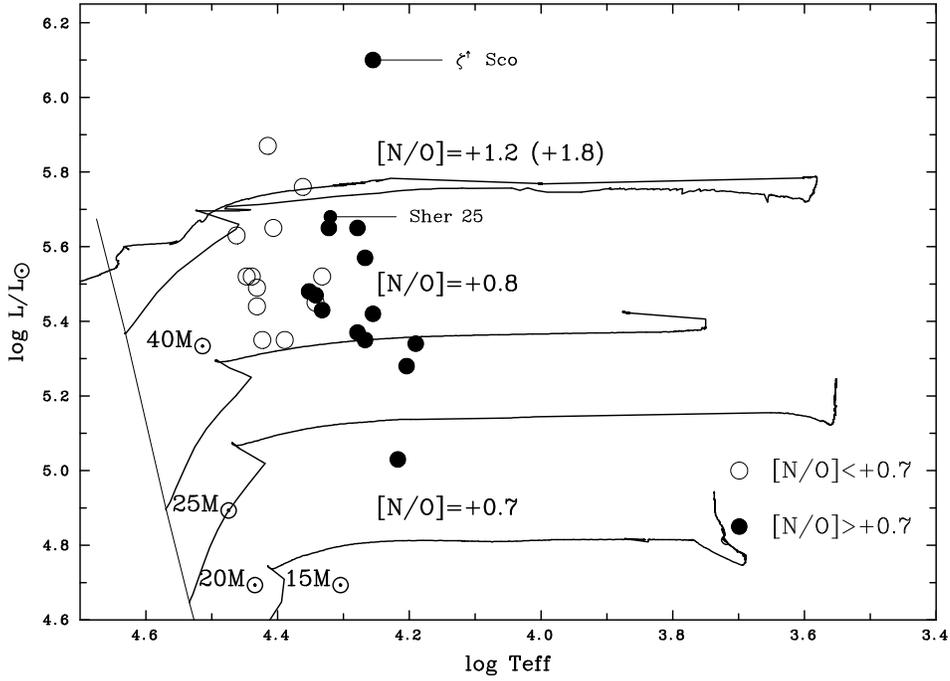}
\end{center} \caption{Observed H-R diagram for our sample of B
supergiants, coded by their degree of N/O enrichment, together with tracks
from Solar metallicity, rotating ($v_{\rm init}$=300 \kms) evolutionary
models (dotted lines) from Meynet \& Maeder (2000) for which predicted N/O
enrichments are indicated for redward (blueward) evolution at \teff\ =
22kK. The location of Sher~25 and HD~152236 ($\zeta^{1}$ Sco) are
indicated.\label{HRD}} \end{figure}

\section{Physical and Wind Properties}

\subsection{Temperature scale}

The inclusion of line blanketing and spherical geometry into O star models
has resulted in a major 15$\pm$5\% downward revision in their \teff\
spectral type calibration (Crowther et al. 2002; Martins et al. 2005).  
For B supergiants, we present our results together with those of Trundle
et al. (2004) and Trundle \& Lennon (2005) for SMC stars studied with
FASTWIND (Puls et al. 2005) in Fig.~\ref{teff}, together with the results
from unblanketed plane parallel models of McErlean et al. (1999). As is
apparent, the revision in B supergiant temperature scale is modest, with
typical revisions of $-$1 to $-$2kK. Note that results for line blanketed
versions of CMFGEN and TLUSTY are consistent to within $\pm$0.5kK for
several stars in common between the present study and Urbaneja (2004).

The location of our stars in the H-R diagram is presented in
Fig.~\ref{HRD}, such that the majority are consistent with initial masses
in the range 20--40 \msun, whilst $\zeta^{1}$ Sco is rather more
massive. Spectroscopically derived masses span a wide range, from
8 up to 76 \msun, based upon surface gravities reliable to \logg\ $\pm$
0.15 dex. In most cases spectroscopic masses are lower than those
estimated from evolutionary models, indicating another example of a `mass
discrepancy', in common with Trundle \& Lennon (2005).

\begin{figure}[htbp] \begin{center}
\includegraphics[width=0.8\columnwidth,clip,angle=0]{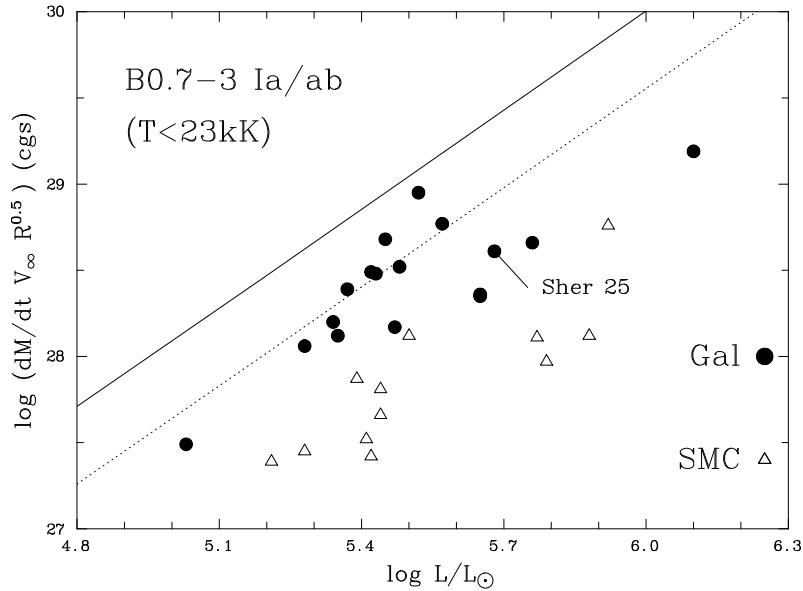} 
\end{center}
\caption{Comparison between the wind momenta of Galactic B0.7--3
supergiants (\teff\ $\leq$23kK)  and studies of SMC counterparts
Trundle et al. 2004; Trundle \& Lennon 2005), assuming 
homogeneous H$\alpha$ mass-loss rates, plus radiatively
driven wind theory predictions by Vink et al. (2000, 2001) for the
Milky Way (solid) and SMC (dotted). Sher~25 is indicated, revealing 
wind properties typical of Galactic B supergiants.} 
\label{wmr}
\end{figure}

\subsection{Mass-loss rates}

H$\alpha$ mass-loss rates compare well with mid-IR excess methods (e.g.
Barlow \& Cohen 1977) and line blanketed FASTWIND models (Urbaneja 2004),
but exceed those derived by Kudritzki et al. (1999) by a factor of three,
who used an unblanketed version of FASTWIND for B2--3 supergiants.
Spectroscopic studies in which clumped winds are included tend to produce
superior spectral fits to far-UV lines (Evans et al. 2004b), suggesting
comparable clumping factors for the inner (H$\alpha$ line) and outer
(mid-IR continuum) forming regions.

In general, wind momenta of Galactic  B0--0.5 subtypes
lie close to the 
Vink et al. (2000) prediction. If winds of early B supergiants are clumped 
in the H$\alpha$ line forming region, we would need to shift the observed 
wind momenta to lower values. Fig.~\ref{wmr} presents wind momenta for 
Galactic B0.7--3 supergiants (\teff\ ($\leq$ 23kK)  together with 
corresponding SMC 
results from Trundle et al. (2004) and Trundle \& Lennon (2005), plus
theoretical predictions from Vink et al. (2000, 2001).
A clear separation between Galactic and SMC supergiants is apparent, with 
approximately the offset predicted by Vink et al. (2001). However, 
measured values lie typically  $\sim$0.5 dex below the Vink et al. 
calibration. As discussed above, if H$\alpha$ 
derived mass-loss rates of OB stars need to be corrected for clumped 
winds, as suggested by recent observational evidence, including Evans et 
al. (2004b) for B supergiants,  the difference between theory  and 
observation would be further exacerbated.

\begin{figure}[ht!] 
\begin{center}
\includegraphics[width=0.8\columnwidth,clip,angle=0]{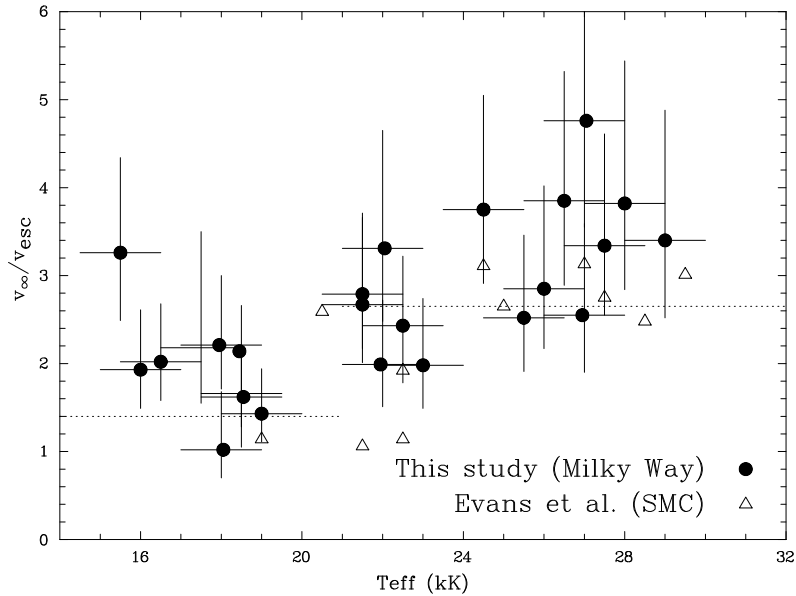}
\end{center} \caption{The ratio $v_{\infty}/v_{\rm esc}$ as a function of
effective temperature for Galactic B supergiants from the
present study (filled circles), together with results from Evans et al.
(2004a, open triangles)  for SMC B supergiants. Error bars relate to
uncertainties in temperature ($\pm$1 kK), gravity ($\pm$0.15 to 0.2 dex)
and absolute magnitude ($\pm$0.3 to 1 mag).\label{bistab}} \end{figure}

\subsection{Bistability jump}

In Fig.~\ref{bistab} we present our current results, together with values
from Evans et al. (2004a), based on measured HST/STIS wind velocities, 
plus
model atmosphere results from Evans et al. (2004b), Trundle et al. (2004),
Trundle \& Lennon (2005). In contrast with the discontinuity 
identified by Lamers et al. (1995), we identify a gradual downward trend  
of \vinf\ / \vesc\  with temperature, albeit with a large 
scatter.   For \teff $>24$kK (approximately B0.5\,Ia and earlier),
\vinf\ / \vesc\ $\sim 3.4$, for 20kK$\leq$ \teff\ $\leq$24kK
(approximately B0.7--1\,Ia)  \vinf\ / \vesc\ $\sim 2.5$, and for
\teff\ $<$20kK (approximately B1.5\,Ia and later) \vinf\ / 
\vesc\ $\sim 1.9$. This reveals that the B1 `jump' is misleading in the
context of normal B supergiants.  Prinja \& Massa (1998) came to similar
conclusions based on a larger B supergiant sample, albeit with an adopted
subtype-temperature calibration. 

Empirical values of the Lamers et al. (1995) `bistability jump' were
adopted by Vink et al. (2000) in their radiatively driven wind
calculations. These predictions have subsequently been used in
evolutionary models (e.g. Meynet \& Maeder 2000).  Consequently, our
re-determination of physical parameters and wind properties of early B
supergiants has potential consequences for evolutionary and spectral
synthesis calculations.

\subsection{Photospheric CNO abundances}

Fig.~\ref{CNO} provides a summary of our derived CNO abundances. With
respect to the current Solar values (Asplund et al. 2004), carbon is
depleted by 0.4$\pm$0.4 dex, oxygen is mildly depleted by up to 0.5 dex,
whilst nitrogen is enriched by 0.6$\pm$0.4 dex. Mean [N/C] and [N/O]
ratios indicate evidence for partial CNO processing within the 
photospheres  of B supergiants. Our results support previous abundance
estimates from the literature, which were typically based upon 
differential abundance analyses from
plane-parallel LTE or non-LTE model atmospheres.

On average, the observed degree of CNO processing agrees well with recent
evolutionary models for 20--40\msun, for which [N/C] = +0.8 to +1.3 and
[N/O] = +0.7 to +1.2 were predicted via rotational mixing ($v_{\rm init}$
= 300 \kms) at $\sim$22kK on the redward evolutionary track (Meynet \& 
Maeder 2001).  Spectroscopically, all our sample of 
supergiants are morphologically normal, except for $\kappa$ 
Cas (HD~2905, BC\,0.7Ia) which indeed possesses the lowest N enrichment in 
our sample with [N/O] = +0.1 dex. 
Conversely, $\zeta^{1}$ Sco (HD~152236, B1.5~Ia$^{+}$), one
of two hypergiants in our sample, possesses
the highest N enrichment of [N/O] = +1.3, that is more typical of nebulae 
associated with Luminous Blue Variables, e.g. [N/O] = +1.3 to +1.9 dex for 
AG Car and  R127 (Lamers et al. 2001).

\section{Sher~25 in NGC~3603}

Sher~25 (B1.5~Iab, Moffat 1983) is associated with a ring nebula 
and apparent bipolar outflows (Brandner et al. 1987a). The ring nebula is 
reminiscent of the inner nebula
associated with SN~1987A, which is presumed to have been ejected from the 
Sk--69$^{\circ}$ 202 
(B3~I, Walborn et al. 1989) progenitor. Both nebulae share physical
sizes of $\sim$0.5 pc, with dynamical ages of several thousand years and 
apparent nitrogen enrichment (Panagia et al. 1996; Brandner et al. 1997b).
Hendry et al. (2006) present studies of the ring nebula and
bipolar lobes from
Sher~25, plus that of the giant HII region NGC~3603 toward 
which Sher~25  is located. 

\begin{figure}[ht!] \begin{center}
\includegraphics[width=1.0\columnwidth,clip,angle=0]{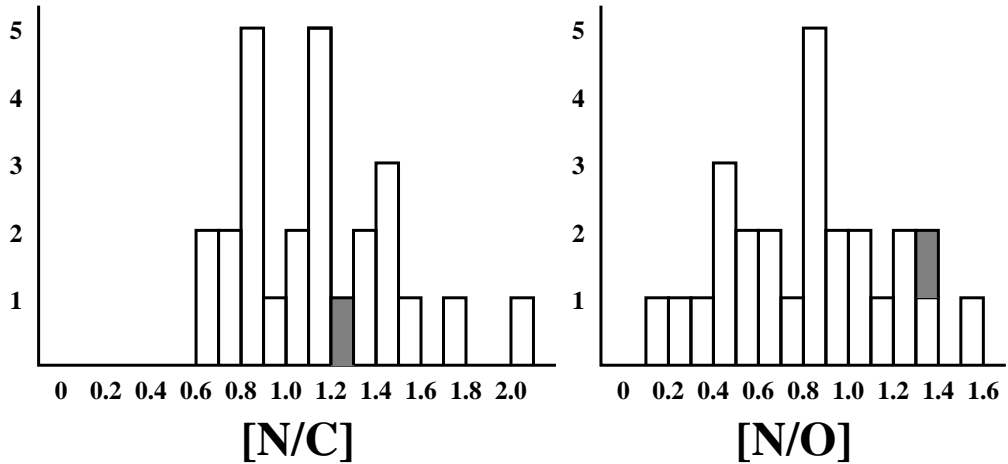} \end{center}
\caption{[N/C] (left) and [N/O] (right) abundances for Milky Way B
supergiants from the present CMFGEN study (relative to Asplund et al.
2004), typically revealing partially CNO processed material, in which
Sher~25 is indicated in grey. Although the relative abundances presented here 
are anticipated to be robust, there remains considerable absolute uncertainty, 
since results for a subset of the present sample using FASTWIND differ by up to 
0.5 dex (Urbaneja 2004).\label{CNO}} \end{figure}

The stellar and wind properties of Sher~25 are typical of other
Galactic early B supergiants in our sample 
(recall Figs.~\ref{HRD}--\ref{wmr}),  as derived from optical 
AAT/UCLES spectroscopy, i.e. \teff\ = 21kK, $\log$ \lstar\ / \lsun\  =  
5.68, \mdot\ =  1.2$\times 10^{-6}$ \msunyr\, \vinf $\approx$ 750  \kms\ 
using a distance of 7.9~kpc (Russeil 2003). 
The surface elemental abundances of Sher~25 derived here,
[N/C]=+1.3 and [N/O]=+1.2, are compared with other B supergiants
in Fig.~\ref{CNO}. Table~\ref{sher25} provides a comparison with
previous photospheric derivations (Smartt et al.
2002) plus ring nebula and bipolar lobe 
abundances in Table~\ref{sher25}.  Note that
our derived [N/O] ratio agrees perfectly with the mean
nebular ratio from Hendry et al. (2006), whilst the giant HII region is 
close to Solar composition (see also Esteban et al. 2005).

If we compare the location of Sher~25 in the H-R diagram with evolutionary
models of Meynet \& Maeder (2000) it lies close to an initial 
40$M_{\odot}$ rotating with $v_{\rm init}$ = 300 \kms\ (recall 
Fig.~\ref{HRD}). If we adjust the predictions from Meynet \& Maeder 
to the Solar abundances from Asplund et al. (2004)  one expects a
nitrogen  enhancement of +1.0 dex and oxygen depletion of --0.2 dex on the
redward track after 5.1~Myr, i.e. [N/O] = +1.2 dex, in perfect agreement 
with the mean nebular and stellar CMFGEN results. For the post-RSG stage 
after 5.2~Myr, [N/O] = +1.8 dex is predicted. Consequently, Sher~25 is 
consistent with a single star evolving towards the RSG phase.

However, amongst the most successful models to explain 
SN~1987A during its B supergiant phase are those of Podsiadlowski (1992), 
involving either accretion from, or merger with, a binary companion
during the common envelope evolution in the RSG stage of Sk--69$^{\circ}$ 
202. Assuming membership of NGC~3603, Sher~25 is
more luminous than Sk--69$^{\circ}$ 202 -- for which $\log$ \lstar / \lsun
$\sim$ 5.1. If Sher~25 has undergone a similar evolution,  which may be 
suggested  from its recent  nebular 
ejection, this star would not be 
expected to possess  CNO abundances typical of normal B supergiants 
evolving towards the RSG phase. The dynamical age of the ring nebula 
and bipolar lobes is $\sim$8700~yr (Brandner et al. 1997b, adjusted for a 
distance of 7.9~kpc).

One final puzzle regarding Sher~25 is that it lies only 20 arcsec 
(1~pc) in projection from the
central ionizing cluster, HD~97950, in NGC~3603, yet its inferred
age of $\sim$5Myr greatly exceeds that of HD~97950 which is $\sim$2Myr 
(Crowther \& Dessart 1998). HD~97950 closely resembles R136 in the LMC 
(Walborn 1973; Moffat 
1983), yet unlike R136, it does not possess an extensive halo of earlier 
phases of massive star formation. Perhaps Sher~25 is a foreground object,
which may plausibly be inferred from the presence of other stars which are 
older than the HD~97950 cluster to one side (Brandner et al. 
1997a). In this case the stellar luminosity and the size of the ring
nebula are smaller than adopted here, although we note its physical
and wind properties are typical of other Galactic early B supergiants.  
Alternatively, close binary  evolution may be responsible for its unusual 
position in the H-R diagram.

\begin{table}
\caption{Photosphere abundances of Sher~25, together with nebular 
abundances for the ejecta nebula and NGC~3603 HII region, 
with reference to Solar abundances of log(O/H)+12=8.66 and 
log(N/H)+12=7.78
from Asplund et al. (2004).}\label{sher25}
\begin{center}
\begin{tabular}{lcccl}
\hline
         & log O/H & log N/H & [N/O] & Reference \\
         &      +12    &        +12 \\
\hline
Star (CMFGEN)& 8.14 & 8.48 & 1.2   & This study \\
Star (TLUSTY)& 8.87 & 8.42 & 0.4   & Smartt et al. 2002\\
Ring nebula  & 8.55 & 8.97 & 1.3   & Hendry et al. 2006\\
Bipolar lobes& 8.64 & 8.88 & 1.1   & Hendry et al. 2006\\
HII region   & 8.56 & 7.47 & --0.1 & Hendry et al 2006\\
\hline
\end{tabular}
\end{center}
\end{table}

\acknowledgements 

Thanks to Maggie Hendry for providing nebular properties of Sher~25 and the
NGC~3603 HII region prior to publication. PAC acknowledges financial support
from the Royal Society.


\end{document}